\documentclass[submission,copyright,creativecommons]{eptcs}
\providecommand{\event}{TICSA - ETAPS 2023} 
\usepackage{graphicx} 
\usepackage{iftex}

\ifpdf
  \usepackage{underscore}         
  \usepackage[T1]{fontenc}        
\else
  \usepackage{breakurl}           
\fi
\title{Product Line Management with Graphical MBSE Views}
\author{Pascal Krapf 
\institute{Syscience\\ Villebon sur Yvette, France}
\email{pascal.krapf@syscience.fr }
\and
S\'ebastien Berthier
\institute{Syscience\\ Villebon sur Yvette, France}
\email{sebastien.berthier@syscience.fr}
\and
Nicole Levy
\institute{CEDRIC-CNAM\\ Paris, France}
\email{nicole.levy@cnam.fr}
}
\def\titlerunning{Product Line Management with Graphical MBSE Views}
\def\authorrunning{Pascal Krapf \& S\'ebastien Berthier \& Nicole Levy}
\begin{document}
\maketitle

\begin{abstract}
Abstract : Reducing the cost and delay and improving quality are major issues for product and software development, especially in the automotive domain. 
Product line engineering is a well-known approach to engineer systems with the aim to reduce costs and development time as well as to improve the product quality. Feature models enable to make logical selection of features and obtain a filtered set of assets that compose the product. We propose to use a color code in feature models to make possible decisions visual in the feature tree. The color code is explained and its use is illustrated. The completeness of the approach is discussed.
\end{abstract}
\begin{abstract}
Keywords : Configuration, variants, product line, model-based system engineering (MBSE)
\end{abstract}

\section{Introduction}
Reducing the cost and delay and improving quality are major issues for product and software development. To achieve these challenges, strategies for reuse and standardization of products and software have been developed. In this way, development and validation of components and software assets are mutualitized over several projects, which reduces the global cost of the product. Indeed, the number of individual assets that are required to build complex products like personal cars, aircrafts, trains or industrial facilities can reach several thousands (from ten to several hundred thousand). Moreover only some of them are present in all products, while the others are associated to particular products. Products can be differentiated by several characteristics: 
\begin{itemize}
\item Products may differ in the offered functionalities. 
\item Products may differ in performance values.
\item Products may differ in the non-functional properties.
\item Products may differ in the chosen execution platform. 
\end{itemize} 
In the automotive domain, a widespread practice is to define from the beginning a Product Line (PL) approach \cite{LePut}. It consists in designing a set of defined products embedding physical and software components developed from a common set of core assets and having a managed set of variable features   \cite{Clements}.

Developing from the beginning a PL, means to focus on the variability and on the potential differences between products. The method we applied focuses on the creation of a product line right from the initial product development stage. The aim is to propose possible variants from the very start, knowing that some others could be added to later. It's a very different approach to parameterization. Defining a parameterized product means concentrating on the common functionalities. The overall architecture is generally not variable and, as a result, non-functional properties are less variable and less emphasized.

Variant management languages and associated tools have been developed with high expressiveness to describe product lines \cite{2007VariantMW, DBLP:conf/splc/Beuche07a, Czarnecki2005, 5070568, Beek2019}. However, up to now, their deployment in the automotive industry is not effective.

In the automotive industry, dozens of development teams are specialized in various domains of engineering. They all contribute to the definition product variants. They each have specific concerns about variability and favor processes and tools suited to their specific concern. But at the end of development, a reduced number of people in the project team has to be able to select the project variants without being expert in all the engineering domains. Thus, the way the variability is structured has to be understandable by people outside the specific engineering domain.

This is why powerful specialized tools are of little help when deploying a PL approach on an industrial project in the automotive domain. 

Our proposal is to define processes methods and an associated tool with simple and visual interfaces made intuitive for users not accustomed to software-oriented tools. Even if less powerful for constraints expression, consistence analysis and solving capabilities than existing ones,
such a tool may be better suited to meet user acceptance in our specific domain.

The first section is the present introduction.

In the second section of this article, we list a set of qualities targeted when configuring a system.

In the third section, we present useful concepts issued from our experience that are frequently used in industries managing different kinds of product lines. 

In the fourth section, we present our approach to build and use PL inspired from FODA's feature models \cite{Kang}. We use for this purpose our existing system engineering modeling platform, which is under improvement.

In the fifth section, we discuss the possibility to handle any kind of logical constraints in the framework we have defined.

\section{Overall strategy and targeted properties}

\subsection{Parameterized software versus Product Line}
The most straightforward way to develop a set of related products is to develop a first product and adapt this product using tune-able parameters and adding components.

The activity of making a system tune-able can be split into: 
\begin{itemize}
    \item identify the possible adjustable parameters and components associated to variabilities, 
    \item define the values to be selected for the parameters and design specific additional components. 
\end{itemize}
Such strategies have been applied by carmakers and have shown some limitations:
\begin{itemize}
    \item Trade-offs are driven by the first designed product, which does not mean global optimization for all products.
    \item Adding variabilities as add-ons considerably increase the complexity of the product and can increase the risk of malfunction.
    \item It generally requires deep knowledge of the product to be able to tune the variable parameters, while companies often want the configuration to be done by non-specialists.
\end{itemize}
 
As a consequence, carmakers are preferring the product line approach \cite{WozniakC15}. A set of defined products that share a common, managed set of features and are developed from a common set of core assets \cite{Clements}. 

\subsection{Qualities targeted when configuring a system}
Following system engineering practices, the first step is to capture needs about the variant configuration management framework. The need is a framework (process and tool) that has the following characteristics \cite{MemCaroline}:
\begin{itemize}
    \item \textbf{Operability}: the number of actions to select a product variant is reduced and available to human decision, both for the first setup and for later updates,

    \item \textbf{Evolutivity}: it is possible to add features and parts to the product line and continue to use former versions. When the product line is enriched, new features are added along with new added constraints,
    \item \textbf{Reusability}: parts and groups of parts can be reused with confidence without modification in new products or new product lines,
    \item \textbf{Simplicity}: no deep knowledge about the system design is necessary to select a variant. Parameterization can be done in a way that is accessible for non-specialists of the domain,
    \item \textbf{Modularity}: Architects can select coherent subsets of the product line by the selection of sets of variants,
    \item \textbf{Consistency}: Compliance with design rules constraining the choice of variants is ensured. These design rules can be of a norm, regulations or chosen method to be followed.
\end{itemize}
 A product line allows to abstract the construction from the configuration of the reusable components, to identify reasoning and decisions behind the selection of a configuration. Reasoning means building a series of relations between causes and consequences while taking into account a set of logical constraints. A PL management framework is likely to satisfy the former list of characteristics.

\section{Field data}
\subsection{Architectural point of views}
Cyber-physical systems are more and more developed using Model based system engineering (MBSE). In these models, systems and their relations with their environment are described by views corresponding to different viewpoints. The approach we were using includes the following viewpoints \cite{IEEE1471}:
\begin{itemize}
    \item \textbf{Operational viewpoint}, focused on the concern of how the system is operated and interacts with surrounding systems.
    \item \textbf{Functional viewpoint} focused on the concern of system functionalities, functional interfaces, functioning modes and behavior.
    \item \textbf{Organic viewpoint} focused on the concern of system components, components allocation of functions and requirements and how internal components interact.
\end{itemize}
These models are used for the product development and are included in the digital twin of the product \cite{Madni}. 
The following paragraphs describe some model elements that are of interest for variant management.
We empathize that variant management is about selecting components (organic viewpoint), but with key drivers coming from the other viewpoints.

\subsection{Operational variability}
A use case is a specific situation in which a product could potentially be used. For example personal cars can be driven on railways, on open roads, in town, on tracks. They can be parked on road, in garden lane or in a garage, etc.
Each use case carries specific requirements the car has to comply with in order to satisfy the customer. Since all customers do not have the same expectations and use cases, different variants of cars are commercialized.
Level of outfitting, seat comfort, acoustics, dynamism, speed, smoothness,  product durability in specific mission profiles are operational characteristics that can be in the scope of variant management. 

\subsection{Functional variability}
Customers can choose among a set of functionalities for their personal car. For example, driver assistance systems, guiding assistance, comfort adaptation, entertainment for passengers, door opening etc. The are an important source of variability. Some of them induce the presence of specific components like sensors and actuators, but others can be activated or deactivated by software. 
End to end functionalities are split into sub-functions forming functional chains. Each sub-function uses inputs to produce some outputs. The combination of these functions ensure the overall product functionalities.

\subsection{Component variability}
\subsubsection{Bill of material}
In industry, the list of all parts or assets that can be purchased to build a product in a product line must be managed. This list is called the "Bill Of Material", or BOM. A 150\% BOM is a list containing the parts to be used to produce the whole set of products of the product line. An individual product will not include all the parts, but only a subset of them: the BOM containing only the parts for an individual product is called 100\% BOM. The BOM is managed as a list in which each standardized part appears only once (an eventual multiplicity will be managed later as we will see).

A BOM may contain up to several hundreds of individual parts, and a large proportion of these parts (frequently 10\% to 50\%) are linked to a variant, as they are not present in all individual products. The first target of variability management is to select efficiently parts corresponding to a specific product. If one wants to select individually each part, one would have to know all the components that are required for each functionality. This choice tends to be impossible for functionalities requiring several hundred thousand parts.  Even for engineers in the appropriate field, this is just impossible.

\subsubsection{Asset library}
Complex systems are often software intensive, meaning that functions are realized by software. Software intensive systems are made of a combination of many interacting software components. Thus, the issue of variant management for physical parts is mirrored in the software domain. Software components are listed in a software library. Assets also include models, specifications, assembly instructions, procedures, tools, validation facilities, safety assessments etc.  Actually, any asset contributing to the product definition can be in the scope of variant management. Thus, physical or purchasable parts listed in a BOM are not enough to define all possibilities to build a product line. It is more flexible and more accurate to consider an asset library that contains any kind of artifacts. 

\subsubsection{Product breakdown structure }
System engineering is the general framework used to develop complex products \cite{IEEE1220, IEEE15288, IEEE1471, EIA632, NASASEH}. A product is broken down into systems. Each system is broken down into subsystems, and so on until reaching individual parts that can be subcontracted and purchased from suppliers. Assets are organized in a tree structure called Product Breakdown Structure (PBS). The PBS contains components that have an active role in the system functioning (sensors, control unit…) as well as components associated to liabilities (tight box, firewall…). Software engineers generally build their software with software components. The software components library is a part of the PBS. Thus, the PBS gives a structured view of the assets that constitute the product.

Engineering teams organize the PBS according to the system breakdown. This breakdown often reflects the organization of the engineering teams and corresponding engineering domains. The manufacturing team has interest to organize the PBS according to how the system is assembled. This may not fit with the engineering team’s organization. The purchasing team may want to structure the PBS according to possible suppliers. The maintenance team may want to organize the PBS in accordance with maintenance schedule and process. If several teams like purchasing, manufacturing, maintenance, use the same PBS then, the structure has to be a compromise between their needs. Companies that try to manage variants by merging all parts in a single PBS that is managed in an Enterprise Resource Planning (ERP) tool often create dissatisfaction in every domain team. This may contribute to the high failure rate of ERP deployment projects \cite{Coskun}.

\subsection{Feature model}
The word “feature” refers to a characteristic or a set of characteristics of the product line. As already discussed, PBS is not the only model element that is impacted by variant management. Thus, there is no reason to have a one-to-one correspondence of nodes of the PBS and features. The PBS structure does not necessarily reflect a selection logic of the components. Requirement satisfaction may involve the contribution of several different parts located in different branches of the PBS. 

General software qualities like cybersecurity, energy consumption efficiency, human machine interfaces, etc. often must be managed with variability. The random selection of software components does not ensure the quality of the final product. 

Companies describe the features that are likely to be variant in a feature model. It describes variable features that can be selected for an individual product within the product line.

Feature Diagrams (FD) are a family of modeling languages used to address the description of products in product lines \cite{Schobbens}. FD were first introduced by Kang as part of the FODA (Feature Oriented Domain Analysis) method back in 1990 \cite{Kang}.

Feature models are generally represented in a tree structure. Each node is a feature that can be selected. A natural rule is to select a son node only if the father node is already selected. In any product line, the feature choice is submitted to rules allowing or forbidding some associations. Rules can result from physics (not enough place), regulation (no such combination of functions), marketing, etc. For example, cars can have diesel, gasoline, hybrid or electrical engines, but only one among this list. These features are not independent, and the dependence is only partially represented by the position in the feature tree. When complex constraints are involved, then it often requires a solver to be able to determine whether a set of selected features comply with these rules. The problem of deciding if a set of logical sentences has possible solutions is known to be NP complex \cite{Schobbens}. 

Variable characteristics usually reach the number of several hundreds to several thousand for cars or aircrafts. Thus, it is still difficult to make choices because of the need to be coherent. Furthermore, the number of possible configurations is still enormous. If 100 nodes can be selected in a feature tree, then the number of possible different products is equal to 2 to the power of 100. This number of combinations cannot be managed extensively. A structured methodology with associated tools is needed.
The domain engineer designs the product line in a way to minimize circular or interwoven constraints. The aim is to make the feature model easily understandable to applications engineers. The PL engineer has to define a smart structure for the product line, and the application engineer needs deep knowledge of the product line to select features without losing time with attempts and errors.

\subsection{Variation criteria}
The PL is described in a model, that contains products assets. Some assets are present in all individual products. These assets form the invariant backbone of the PL. The other assets are present or absent depending on the features that are chosen. A variation criterion is a logical formula, that defines the asset variability. This logical formula is expressed using the features of the feature model. The asset is present in the product if the formula is evaluated TRUE. In this way, assets can be filtered according to the feature selection. The completeness and coherency of this association between assets and features fully relies on the PL design engineer. 

\section{Framework for variant management}
In order to be efficient, companies need a framework of combined and coherent processes, methods and tools. In this section, we describe the framework we have developed to manage a PL. 
Our proposal allows a very broad acceptance of the notion of PL among the multiple actors involved. We have drawn inspiration from a number of existing proposals \cite{BenavidesCTS06, CzarneckiAKLP05}.

\subsection{Processes}
The product line strategy relies on the following major processes: Build the PL, configure a product in the PL and maintain \& enrich the PL.

\subsubsection{Build the PL}
Companies want to have competitive advantages, and to answer more and more customer needs with individual adaptation. Before designing a system, system engineers have to analyze needs. They examine the system’s environment and identify interactions, constraints and available resources. The capture of stakeholder’s needs is the key to system engineering. It is also the first source of variants. Thus, in a PL process, the outcome is not only a set of elicited needs, but also a variability assessment of the PL. Needs capture and analysis is combined with the analysis of the PL variability.

Building the PL includes:
\begin{itemize}
    \item defining a set of assets (components, software, models…) that are designed with the target of addressing a wide range of user needs. 
    \item Building a feature model that describes product features and constraints between them.
    \item Associating assets to features.
\end{itemize}
Assets are associated to features with the target of ensuring modularity and enabling evolutivity criteria mentioned in section 2.

\subsubsection{Configure a PL}
Products to be sold to customers are built as configurations of the product line. The PL contains all assets describing possible products. When a product engineer selects a product for a customer, he defines the features of this specific product. Features are selected in the feature model of the product line. Assets of the specific product are obtained as a consequence of the features. PL assets are filtered according to chosen features to obtain the asset lists of the specific product. Thus, the specific product is a configuration of the product line. This process accounts for the reusability criterion mentioned in section 2.

\subsubsection{Maintain \& enrich the PL}
When engineers have to design a solution for new customer needs or new project requests, they first try to integrate in their design existing assets from the product line. And this shall be done on the system as a whole, considering needs that shall be satisfied, and at the component level for component functionalities and tested qualities. Thus, the design method consists in searching within the existing assets which ones could be reused as they are, which ones could be reused with only small modifications or additional tests and which ones could be integrated in the product via the adaptation of some interfaces or the use of adaptation parts (brackets, connectors, embedding…). 

New assets are developed only if existing assets do not allow to answer the new elicited needs. And if so, they are designed in a way to enable their reuse for future products. Each time a new component is developed, it can be included in the asset library. Asset’s characteristics are standardized and recorded in the aim of reuse. 

The product line also has to be maintained, meaning obsolescence of PL assets is monitored and new assets are developed in the right schedule to replace the obsolete ones without shortage.
This process accounts for the evolutivity criterion mentioned in section 2.

\subsection{Method}
The method we present is intended to describe the product line management and configuration for people who are not necessarily familiar with software development tools. To do so, the steps of the product line use are made visual with graphical diagrams and simple colors codes.

\subsubsection{Association between assets and features}
Assets have to be associated to features in order to model the transition from the asset library to the selected assets constituting a specific product. For each asset, a logical statement defines its presence. This statement uses logical connectors and features. In that way it is possible to use a feature configuration to filter the asset library and obtain the assets of a specific product. The introduction of the feature conditions in the description of the behavior of components allows for a configurable behavioral model of the product line. The granularity of the features and the association to groups of assets in the library strongly influences the number of operations to be done to configure a product. It has a major impact on the operability criterion introduced in section 2. To define a product, it will not be necessary to select individual parts but product features. If features are well structured, then their choice is operable by humans. As an illustration, a limited set of feature choice is proposed to a customer purchasing a car, impacting the presence of dozens of components in the product.

\subsubsection{Constraints expressions in a feature diagram}
Each node in the feature tree represents a decision. At each decision step, constraints are limiting the number of possible choices. In many cases, the constraints concern neighbor nodes. Thus, being able to display in an intuitive way these constraints is of interest. The color of a node can be used to display these constraints. In our method we propose to consider tree type of choices that are displayed by a color code. The color code can be replaced by any other graphical characteristic of the boxes, especially if accessibility for color-blind users is required.

\begin{figure}
 \centering 
\fbox{\includegraphics[width=0.9\textwidth]{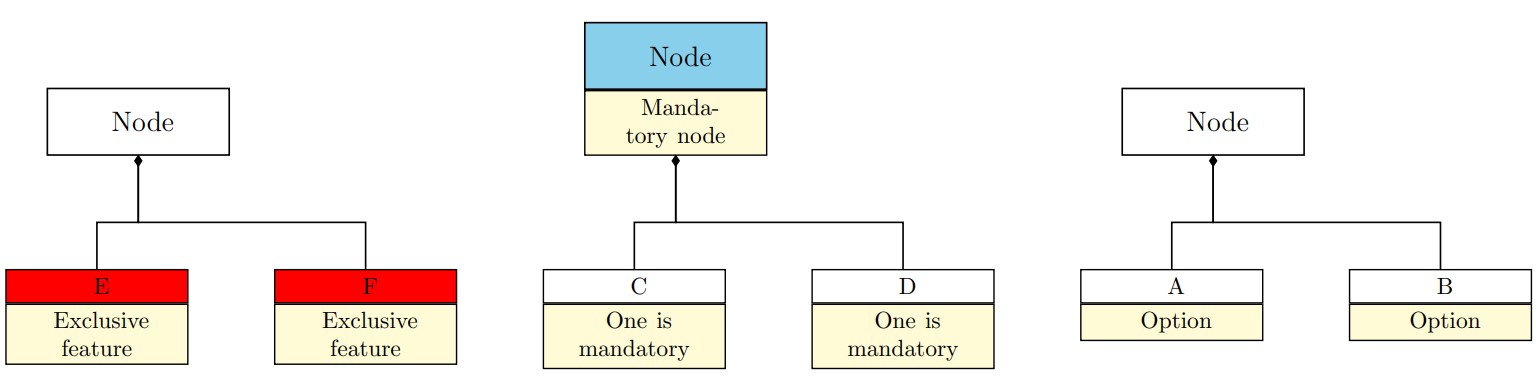}}
      \caption{Representation of the different kinds of choice in the feature model}
\label{fig:Fig3}
      \end{figure}

Optional features are features that can be selected or not without further constraint. Those features are represented in white boxes as shown in Figure~\ref{fig:Fig3}. A white box can be selected or discarded independently of neighbor boxes. If the parent node is selected, then the children selection can be a single child, both of them or none.

Blue is used as shown in Figure~\ref{fig:Fig3} to indicate a mandatory choice: the parent (blue) node has to be kept and at least one of the boxes below has to be selected. When using this colour code, it was found more intuitive for non-specialists to have the blue colour on the upper node, where the decision is taken, rather than on the lower node.

Red is used as shown in Figure~\ref{fig:Fig3} to indicate exclusive options: only one of the neighboring red boxes can be selected. If one is selected, then the neighboring red boxes have to be discarded. The red color is applied on lower nodes. It is a difference with the blue color and was found more intuitive for non-specialist users who have to define a product within the PL. Plus, it allows to combine easily red and blue nodes.

\subsubsection{Product configuration}
\begin{figure}
      \centering 
\fbox{
\includegraphics[width=0.9\textwidth] {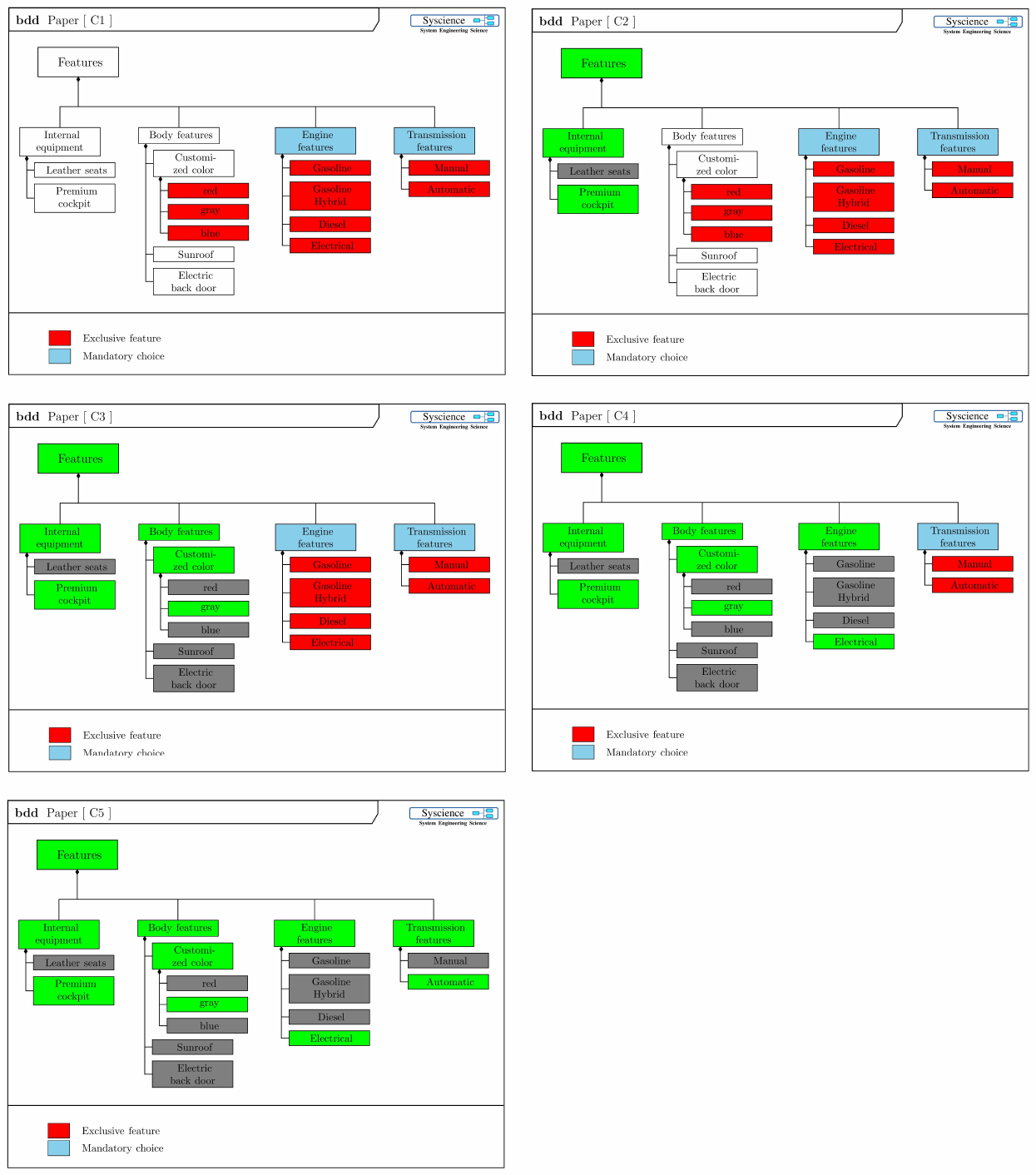} }
      \caption{Representation of successive decisions in the feature model}
\label{fig:Fig6}
      \end{figure}

The most natural way to fill a variant tree is top down. One begins with the upper node and goes along the branches down to the leaves. At each step, the possible choices are defined by the color of the surrounding boxes. The variant selection process is made visual as displayed in Figure~\ref{fig:Fig6} and intuitive to users that have to do it.

A product configuration is obtained by the selection of a set of features that drive the selection of the associated assets. The method to obtain such a configuration requires a set of decisions, whether to keep or not each feature. A feature model is a set of possible decisions, containing also mandatory features. It is important to have them as they can imply sub-decisions. Our feature model defines variable features and constraints between them. In a product configuration, some features are selected, and others are discarded. We use the green color to indicate that a feature is selected, and the gray color to specify that a feature is discarded from a specific product. Thus, a feature model fully colored with green and gray like the one displayed at the bottom of in Figure~\ref{fig:Fig6} is a description of an individual product of the product line. 

Defining a product configuration and verifying at the end if compatibility rules are satisfied is likely to produce configurations incompatible with the rules. The selection has to be organized in a succession of decisions. To carry out this selection, the feature model is scanned down from the root to the leaves. When a node is discarded, then the whole branch below the node is discarded as well. So there cannot be an alternation of green an gray colors down a branch. Coherency rules are checked at each decision step. If a decision would lead to inconsistent feature selection, then the corresponding choice is not possible. Figure~\ref{fig:Fig6} illustrates successive decision steps that lead to a configuration.

Feature selection gives a progressive coloring of the feature model with green and gray. The selection process can be interrupted at any moment. Partially selected feature model can be produced, in which some parts are selected, some others are discarded and some others are still open options. The color of the boxes provides a comprehensive way to describe partial configuration and to define rules to continue the selection process. 

\subsection{Tool}
As the processes to build a new product rely on system engineering, it is natural to use system engineering tools for PL engineering. Thus, variant management shall be embedded in the system engineering tool. One key success factor is to make people working in different domain understand each other and communicate efficiently. Therefore, it is crucial to provide graphical views intended to be intuitive for non-specialists and so to fulfill the simplicity criterion in section 2. We have developed a private model-based system engineering tool that is already in use \cite{LM1, LM2, LM3, LM4}. In a true digital twin perspective behavioral models can be amended by the feature selection. A model of the feature selection is embedded in the tool. In this perspective, we obtain a model of the product line that describes all the products in the product line with their individual characteristics and behavior. While graphical views presented below are available in the tool, the interpretation of constraints stated as logical formulas is still under development.

\section{Discussion}

In the former paragraph, we have proposed a description for a feature model.
Its semantics is similar to the one proposed by \cite{Schobbens}. As different boxes may carry the same label, it describes a Directed Acyclic Graph (DAG). In addition, the colors are used to express “require” and “exclude” relations. Let's take a closer look at how these constraints are used.

      \begin{figure}
      \centering 
\fbox{\includegraphics[width=0.45\textwidth]{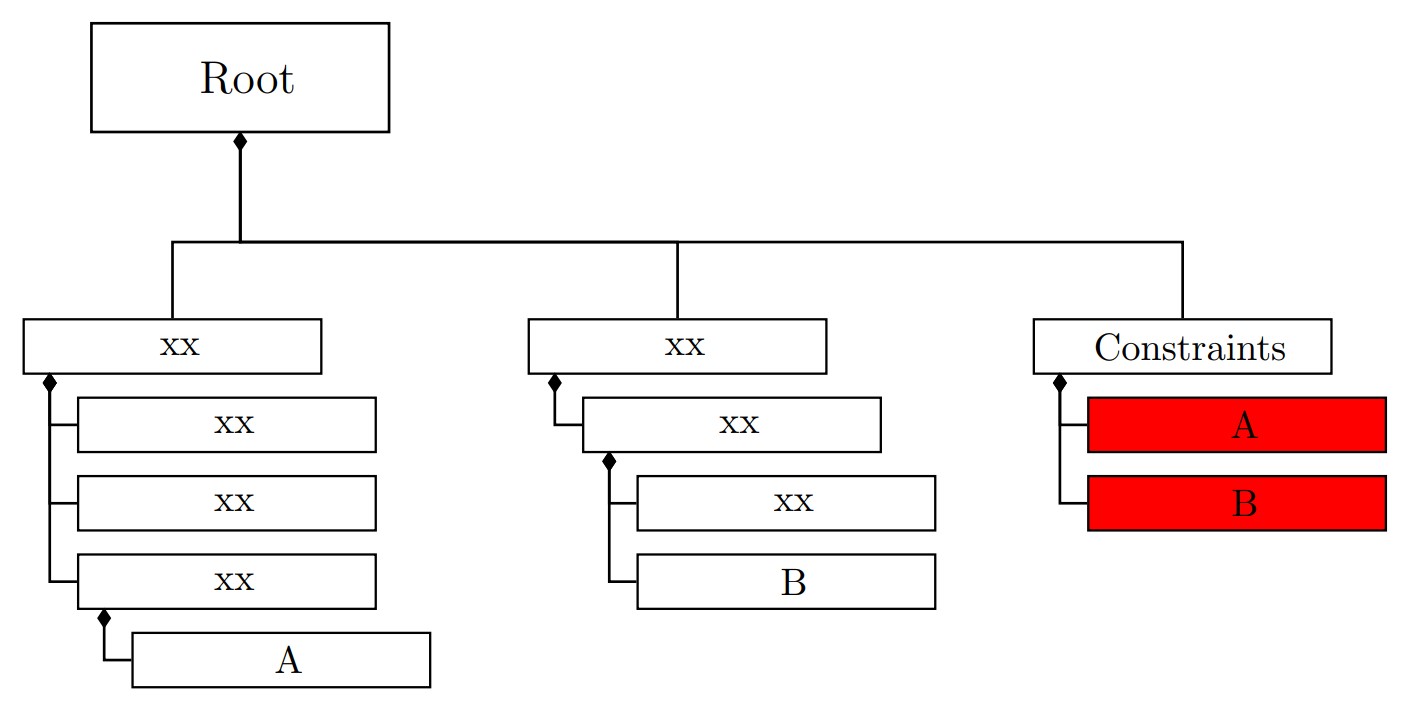}\includegraphics[width=0.45\textwidth]{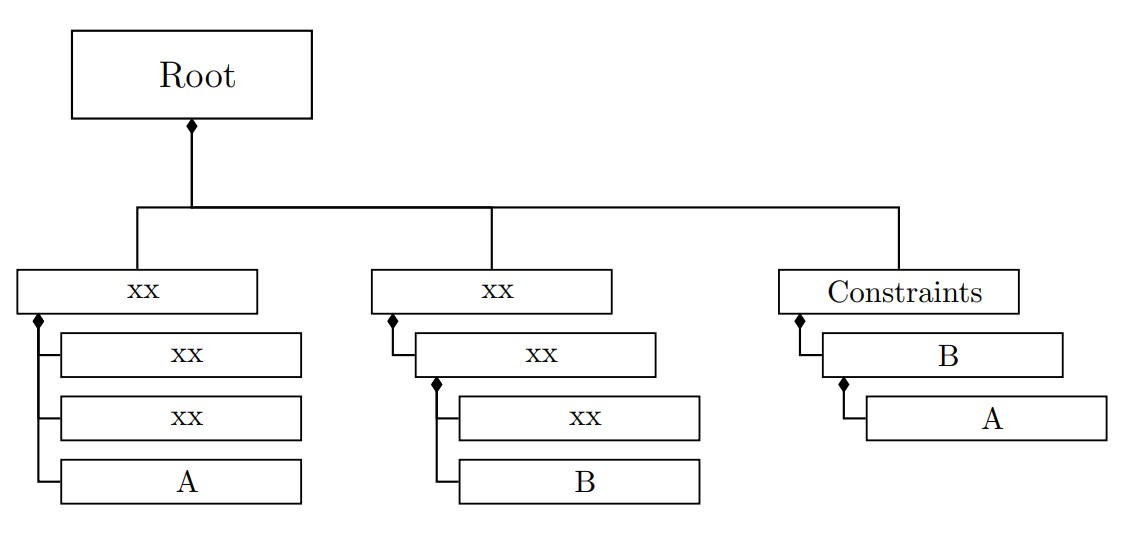}}
      \caption{Representation of the constraints A excludes B and A requires B when A and B are in different branches}
\label{fig:Fig7}
      \end{figure} 
It is clear that the “exclude” relation between neighbor nodes can be directly expressed with the color code. Let A and B be two features in different branches of a variant model. The sentence “Feature A excludes Feature B” means that if feature A is selected, then feature B cannot be selected (let us note that A excludes B is equal to B excludes A). Figure~\ref{fig:Fig7} shows how this constraint can be expressed in our colored box language. Beside the main feature tree, we introduce a new branch labeled “constraints” with node A and B within red boxes, meaning the user has to make an exclusive choice. If the user selects the first A node, then through a decision propagation, the other node labeled A is automatically selected because it has the same label. Node B is automatically discarded because of mutual exclusion with A. 

The sentence “Feature A requires Feature B” means that if feature A is selected, then feature B is mandatory. Figure~\ref{fig:Fig7} shows how this constraint can be expressed in our colored box language. We introduce a new branch labeled “constraints” with a node labeled B. Below this node, a single node labeled A is placed. When the user selects A somewhere in the tree, then decision propagation selects automatically all nodes labeled A. B is automatically selected as a parent of A. 

Thus, our graphical language is able to describe the “require” and the “exclude” relations.

\begin{figure}
      \centering 
\fbox{\includegraphics[width=0.3\textwidth]{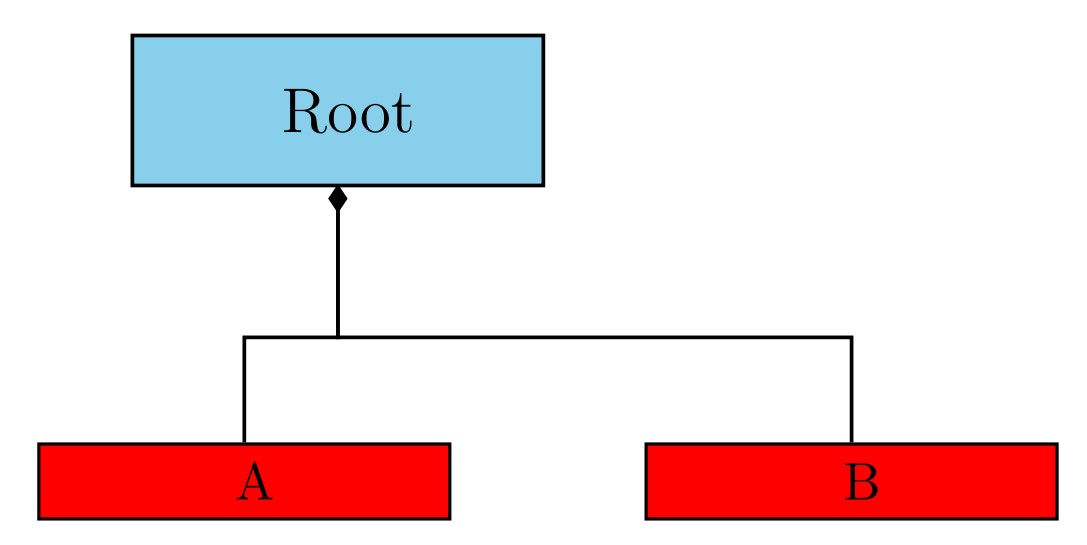} \includegraphics[width=0.3\textwidth]{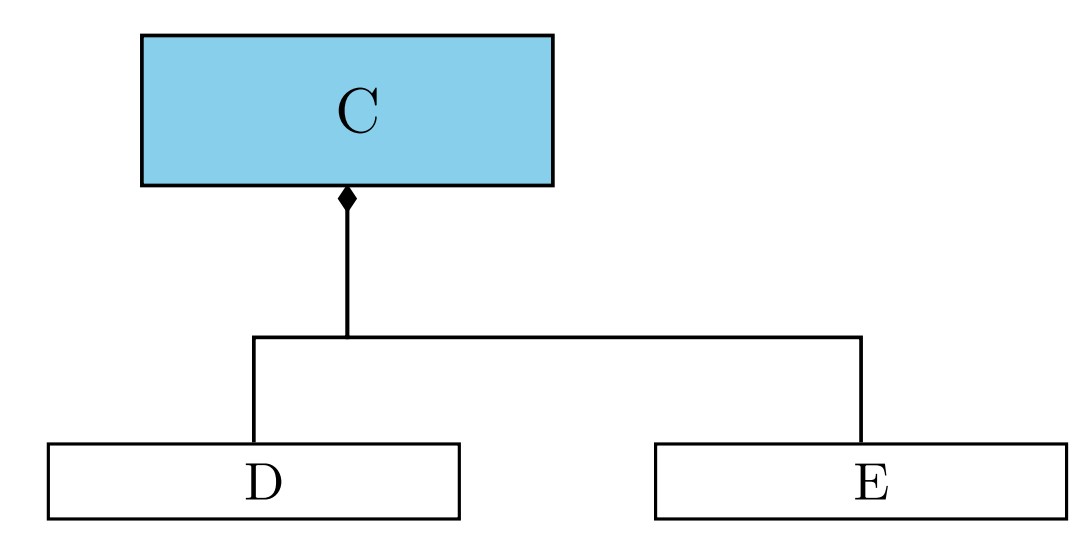} \includegraphics[width=0.3\textwidth]{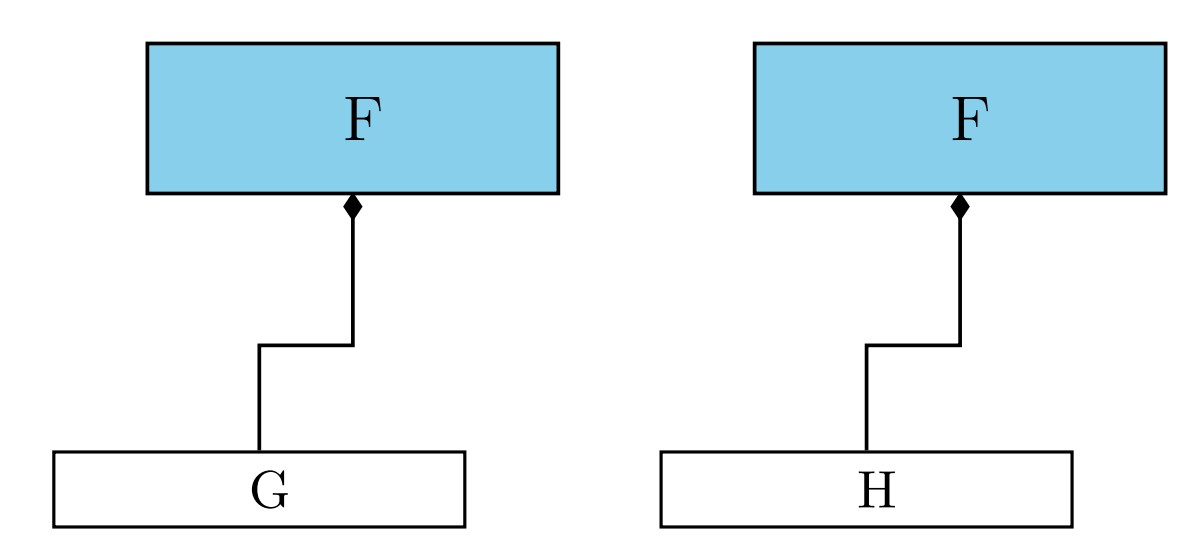}}
      \caption{Representation of the logical constraints A is NOT(B), C implies (D OR E) and F implies (G AND H)}
\label{fig:Fig9}
      \end{figure}

Let us express basic logical constraints with this language. Fig. 9 defines a variable B, that corresponds to NOT(A), a variable C that corresponds to (D OR E), and a variable F that corresponds to (G AND H). Since (AND, OR, NOT) is a complete set of connectors in Boolean logic, any Boolean formula can be expressed by this language. For example a nested constraint ((A AND B) => C) has first to be written as a normal disjunctive formula. After that, it is possible to express it with the colour code. Figures~\ref{fig:Fig9}, gives a general pattern that makes it possible to transform a Boolean expression into a graph with our color definition. The expressiveness of the defined language is universal as soon as we consider feature models that are directed graphs rather than trees. 

Circular constraints are not allowed and should be detected by the tool. When the product line is well defined, the selection of an individual product is straightforward.
By doing so, the consistency criterion mentioned in section 2 is based on the decision propagation between nodes. If a choice leads to a dead-end where any choice left would violate a constraint, then it is necessary to backtrack and undo the last choice. It can be a future improvement to analyze the structure of constraints and disable any choice leading to a dead-end.

\section{Conclusion}

In this paper we have discussed the product line approach as a way to design products that can be configured according to customer needs. 
The product line approach is suited for the automotive industry. However the large number of actors involved in the definition of product variants and product configuration is limiting the use of complex tools. We have therefore defined a framework for product line management in order to address this issue. The proposed method is based on a color code that makes possible decisions visual and intuitive for users unfamiliar with variant management. The method has been illustrated with an example. Our model-based system engineering tool was used to draw the diagrams. The completeness of the method was discussed. 

The current development of our tool includes an allowed configuration only if logical constraints are satisfied at each decision step. In that way, a correct product line feature model does not require a solver to check coherence while configuring a product, as constraints are taken into account at each decision step. 
The scaling to larger product lines relies on a well structured feature model, broken down into as many sub-trees as necessary to keep each graphical view understandable.

We plan to apply our approach of introducing a product line when defining an initial requirement for a system in an industrial domain, using the platform under development.

\section*{Acknowledgment}
      The authors are grateful to Caroline CABY for her insights and the very fruitful discussions she was involved in.

\end{document}